\documentclass{iucrjournals}

\usepackage{minted}
\usepackage[version=4]{mhchem}
\usepackage{lineno}
\nolinenumbers

\title{PowderLine: a programmatic powder diffraction analysis application}
 
\author[a]{Adam A. Corrao\IUCrOrcidlink{0000-0001-6111-8959}}
\author[b]{Jennifer A. Perez\IUCrOrcidlink{0009-0001-7770-9363}}
\author[b]{John D. Langhout\IUCrOrcidlink{0000-0002-0945-2113}}
\author[b]{Megan M. Butala\IUCrOrcidlink{0000-0002-7759-5300}}
\author[a]{Thomas A. Caswell\IUCrOrcidlink{0000-0003-4692-608X}}
\author[a]{Daniel Olds\IUCrEmaillink{dolds@bnl.gov}\IUCrOrcidlink{0000-0002-4611-4113}}%

\affil[a]{National Synchrotron Light Source II, Brookhaven National Laboratory, Upton, NY 11973, USA}%
\affil[b]{Materials Science and Engineering, University of Florida, Gainesville, FL 32611, USA}%
\begin{document} 
\maketitle

\section{Introduction}
Powder diffraction is a foundational materials characterization method that is used routinely in academic, government, and industrial settings across practically all fields, including energy storage, catalysis, pharmaceuticals, ceramics, semiconductors, polymers, mining, minerals, cultural heritage, and cosmetics \cite{Kaduk2021}. While it is often used as a material fingerprinting technique through phase identification and pattern matching tools, there is rich structural, chemical, and morphological information embedded in the positions, intensities, and shapes (\textit{i.e.,} profiles) of diffraction peaks. This includes the unit cell description and lattice parameters, atomic positions, site occupancies, chemical substitution and disorder, relative phase fractions, and microstructural details like crystallite size and shape, lattice microstrain, and stacking faults. Determining and quantifying the features present is the central goal of quantitative powder diffraction analysis.

Extracting this quantitative information typically requires modeling the full diffraction pattern, most commonly through Rietveld refinement \cite{Rietveld1969, Young1993}. Whole-pattern fitting methods are mature and capable, but obtaining reliable results can require considerable expertise. A given fits refinement strategy is often informed by a mixture of experience, chemical intuition, and trial-and-error. Such approaches are a poor match for high-throughput experiments and autonomous self-driving laboratories, which can generate volumes of data at rates incompatible with conventional hand driven analysis approaches \cite{Szymanski2023, Yotsumoto2024, Corrao2025}. 

Machine learning (ML) and artificial intelligence (AI) have been applied to powder diffraction analysis with considerable success on specific tasks, including phase identification and classification \cite{Maffettone2021, Szymanski2024, Ozaki2025}, phase quantification and microstructure estimation from known structures and simulated patterns \cite{Dong2021, Souesme2026}, and crystal structure determination with generative models \cite{Riesel2024, Li2025, Li2026b}. However, these models are typically trained for narrow, well-defined tasks, and they do not yet generalize to the diversity of materials encountered in practice \cite{Szymanski2021, Leeman2024, Rincon2025}. For example, a measured pattern may correspond to a material that is not yet present in structural databases, distinct structures can produce closely similar patterns, and a single structure type can span a wide range of compositions and chemistries. Real samples add further complexity through the nearly unlimited space of secondary phases, phase mixtures, and intricate features such as chemical order/disorder and anisotropic peak broadening, all of which affect diffraction peak intensities, positions, and shapes \cite{Yanxon2023, Yadav2026}. These factors make it difficult to train a model that generalizes across phase spaces, crystal systems, and the diversity of features in diffraction data \cite{Schuetzke2021}. Traditional whole pattern fitting methods remain essential for reliable quantitative analysis, and automating such methods is what such emerging workflows require \cite{Mun2026}.

Several programmatic and automated approaches to Rietveld refinement have been developed to address these needs. These include control-file driven programs (\textit{e.g.,} TOPAS, FullProf, and BGMN) \cite{Coelho2018, RodriguezCarvajal1993, Doebelin2015}, scriptable interfaces to established refinement programs (\textit{e.g.,} GSAS-II) \cite{Toby2013, ODonnell2018}, automation layers that wrap these programs for batch and parallel processing (\textit{e.g.,} MILK and Spotlight) \cite{Savage2023, Biwer2025}, and pipelines that couple refinement to automated phase identification (\textit{e.g.,} DARA) \cite{Fei2026}. While these approaches effectively serve the workflows they were built for, they do not establish a generalized approach to how a refinement is specified and executed. To the best of our knowledge, no powder diffraction analysis software currently provides a single validated, machine-writable description of a refinement that can serve as common input for people and automated workflows alike.

We developed PowderLine to close this gap and provide a composable, extensible, and scalable approach to powder diffraction analysis. PowderLine encapsulates a complete refinement into a single recipe written in JSON. PowderLine can validate the recipe against a versioned schema, execute it through supported software (currently supported includes GSAS-II and TOPAS), and return the refinement results as structured Python objects. This enables controlled and reproducible runs whether refinements are hand-executed by humans, run as part of an automated workflow, or orchestrated by AI agents. In this paper, we describe the design of PowderLine and its recipe schema, demonstrate its use for a multi-phase refinement on a disordered rocksalt cathode materials, and discuss how PowderLine enables future applications.

\section{Architecture and design of PowderLine}
\subsection{Design philosophy and overview}
\begin{figure}[t] %
\begin{center}
\includegraphics[width=0.5\textwidth]{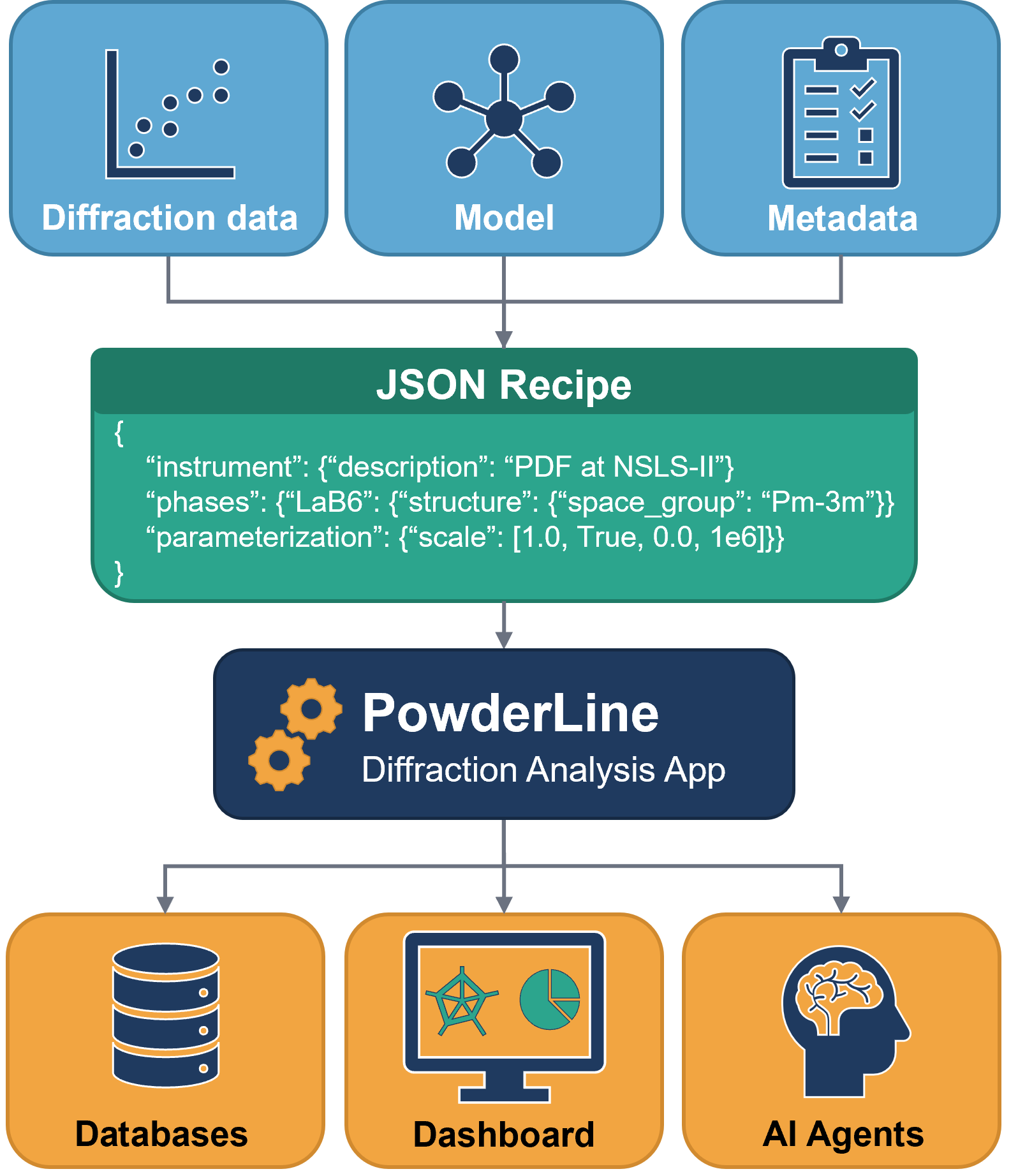} 
\end{center}
\caption{PowderLine workflow diagram that shows the encapsulation of a refinement within a JSON recipe.} 
\label{fig:figure1}
\end{figure}

PowderLine was developed around the core idea that a complete refinement should be fully encapsulated in a single declarative recipe that conforms to a versioned schema. The three main parts of a recipe are the diffraction data, the refinement model, and metadata (Figure \ref{fig:figure1}). Metadata required by PowderLine are the specific schema and the schema version. Beyond these requirements, Powderline supports additional user-defined metadata that can be generated upstream of analysis such as refinement intent, sample provenance, measurement conditions, and the recipe generation method. The result is a metadata-rich dataset suitable for downstream statistical analysis and interpretation. By version controlling the schema, refinement models can evolve as new features are added to PowderLine. All schemas are fully documented and versioned, making PowderLine extensible and refinements reproducible. 

Powderline is an application that interfaces with and drives refinement engines, providing programmatic use of the existing capabilities in analysis software. It composes directly with the scientific Python ecosystem (\textit{e.g.,} NumPy, SciPy, pandas), keeps data flow in memory, and both ingests and returns structured data types. This makes PowderLine easy to integrate with databases, analysis dashboards, and AI agents. A single recipe fully specifies a refinement, and the same recipe produces the same result whether it is executed by a person, as part of a script, a workflow, or by an AI agent.

\subsection{Recipe schema}

\begin{figure}[ht] %
\begin{center}
\includegraphics[width=0.9\textwidth]{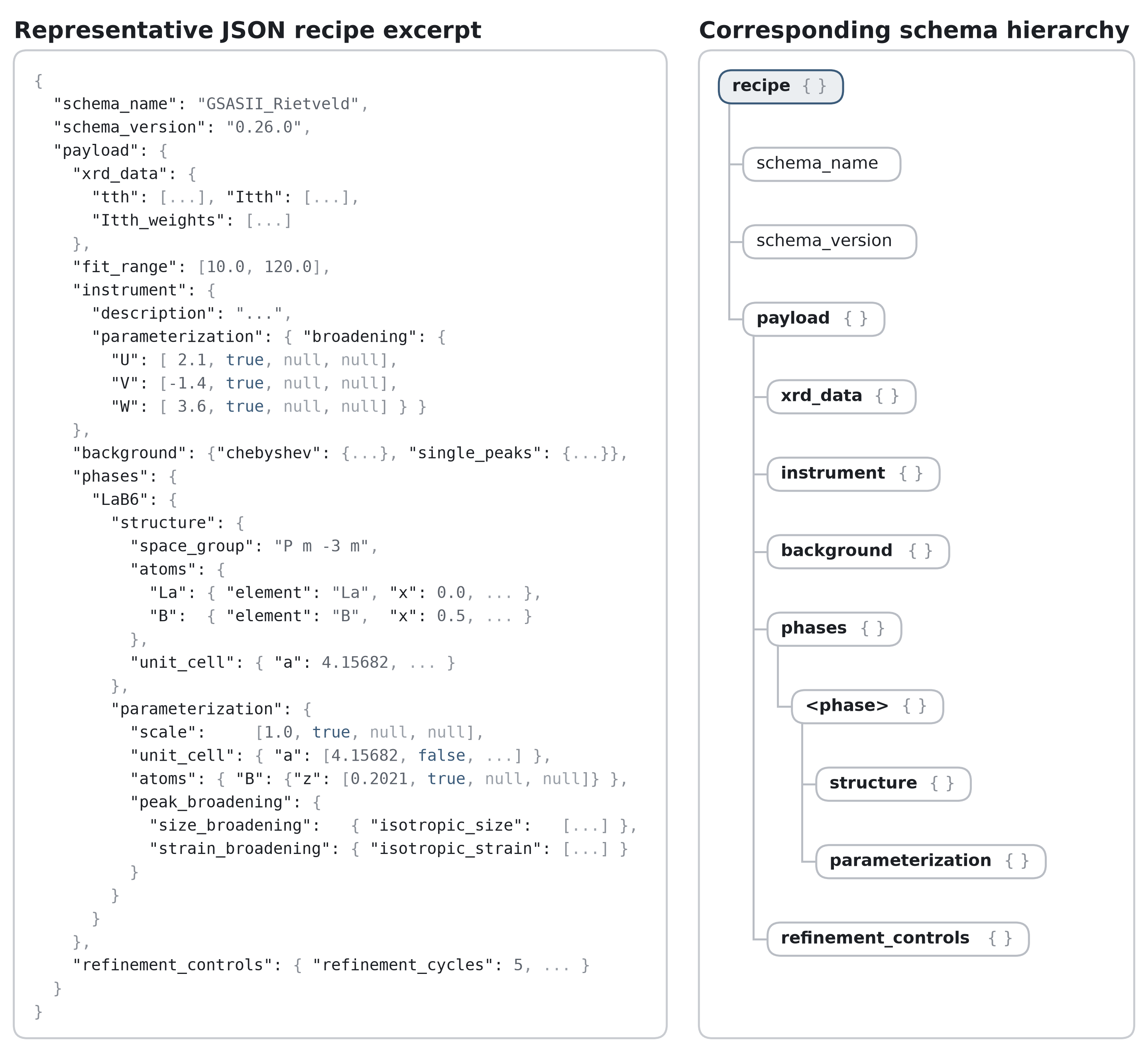}
\end{center}
\caption{Left: Example JSON recipe for fitting powder diffraction data collected on \ce{LaB6} (NIST SRM 660c, $Pm\bar{3}m$, \textit{a} = 4.15682(6)) to determine the instrument profile contribution. Right: corresponding schema hierarchy that the example recipe follows.} 
\label{fig:figure2}
\end{figure}

A PowderLine recipe is a single JSON document with a hierarchical structure (see Figure \ref{fig:figure2}). At the top level a \texttt{schema\_name} specifies the workflow (Rietveld refinement or single-peak fitting), a \texttt{schema\_version} ties the recipe to a validated schema, and a \texttt{payload} expresses the refinement relative to the schema and version. 

The payload organizes the refinement into named blocks that correspond to the familiar ingredients of a whole-pattern fit such as the phases and their parameterization, background model(s), and the refinement controls. The payload also includes components that are typically file-bound such as the diffraction data, instrument details (\textit{e.g.,} instrument profile function), and crystal structure info. Rather than a sequence of software-specific commands, the recipe is a structured description of the analysis. An excerpt from an example recipe for fitting diffraction data on a LaB\textsubscript{6} standard reference material is shown in Figure \ref{fig:figure2}, demonstrating how a real refinement is specified (left) and how this corresponds to the hierarchical schema (right). 
Refinable parameters are represented throughout the schema in a single standard form, the Pydantic \texttt{RefinementParameterModel}, which pairs a value with a refine flag and optional bounds. The bracketed \texttt{[value, refine\_flag, ...]} fields in Figure \ref{fig:figure2} show this in use for several parameters, such as the Gaussian instrumental broadening parameters (U, V, W), the \ce{LaB6} phase scale factor, and boron's non-special \textit{z} atomic coordinate. This approach provides a consistent way to state which parameters are fixed and which are refined. Refine flags are set per parameter, and when needed PowderLine implements the constraint logic required to reflect that. For example, a user can simply specify the refinement of an individual lattice parameter (\textit{e.g.,} hold \textit{a}, refine \textit{c} for a tetragonal cell) or atomic coordinate directly.

\subsection{Validation}
PowderLine validates every recipe against its schema before a refinement runs, confirming that the fields required for the chosen workflow are present and that all fields in the recipe are well formed. For example, a Rietveld recipe must define at least one phase and the instrument profile function, while a single-peak fitting recipe must not contain phases. Individual values provided through the \texttt{RefinementParameterModel} are also checked against parameter-specific constraints that we impose, such as positive unit cell lengths and cell angles between 0 and 180 degrees. When a recipe violates one of these rules, validation fails before any computation and returns a specific message that identifies the offending field and the reason (\textit{e.g.,} size or strain broadening defined without specifying a model).

To that end, every recipe must state the version it was written against: a recipe whose version PowderLine does not support is rejected outright. This ensures the intent of a recipe is unambiguous and reproducible as the schema evolves. Validation can also be done separately from a refinement run, allowing a recipe to be checked before it is submitted for computation. This is particularly useful when recipes are generated programmatically. 

\subsection{Execution modes}
Once validated, a recipe is executed by constructing the refinement from the recipe's contents. This involves a translation layer between the recipe and the specific refinement software's requirements. The user prepares no separate input files and parses no output. When on-disk resources are required by the refinement program (\textit{e.g.,} project files for GSAS-II, input files for TOPAS), PowderLine manages temporary files and reads results back directly. The files can optionally be saved if a user wished to retrieve the refinement artifacts for inspection or debugging.

Two execution modes are currently available through PowderLine: a direct mode that loads the refinement software for each run, and a persistent server mode via FastAPI that keeps the refinement software alive across many refinements. Each recipe is self-contained, so refinements are independent of one another and can be run in parallel, which supports scalable, high-throughput analysis. 

\subsection{Outputs}
A completed refinement returns its results as standard Python variables such as structured tables, scalars, and arrays. These results include the fit profile as well as all refined parameters with estimated standard deviations (ESDs). In this way, descriptors of a refinement are immediately available for further analysis and interpretation. This decoupling of analysis intent from the inner workings of the specific software means that the refinement package does not impose its unique file formats or input/output handling on the rest of the workflow. As a result, interactive sessions, scripts, and automated workflows can consume refinement results identically, removing the typical friction encountered when integrating analysis software.

\section{Worked example: a disordered rocksalt cathode material}
We demonstrate PowderLine for the analysis of synchrotron X-ray powder diffraction data from a disordered rocksalt (DRX) cathode material with a nominal composition of Li$_{1.2}$Mn$_{0.4}$Mg$_{0.2}$W$_{0.2}$O$_2$ that was measured in a previous study \cite{Langhout2026} on the Pair-Distribution Function (PDF, 28-ID-1) beamline at the National Synchrotron Light Source II (NSLS-II). The data was measured in a 1 mm Kapton\textsuperscript{\textregistered}\ capillary under ambient conditions under typical beamline operation ($\lambda$ = 0.1665 \AA). The sample contains two crystalline phases, the DRX phase of interest (face-centered cubic, $Fm\bar{3}m$) and a secondary Li$_4$MgWO$_6$ phase (monoclinic, $C2/m$) formed during solid-state synthesis.

The complete two-phase refinement is expressed as a single recipe. This includes the measured data, the instrument profile function, a six-term Chebyshev background, each of the phases structural models, and the parameters to be refined. We chose to refine for each phase its scale factor, lattice parameters, and strain broadening, while the instrument parameters and the atomic structure (positions, occupancies, and displacement parameters) were fixed (full recipe provided in SI). The refinement was carried out with a single call to PowderLine, which validated the recipe, ran the refinement, and returned the results.

\begin{figure}[ht] %
\begin{center}
\includegraphics[width=0.9\textwidth]{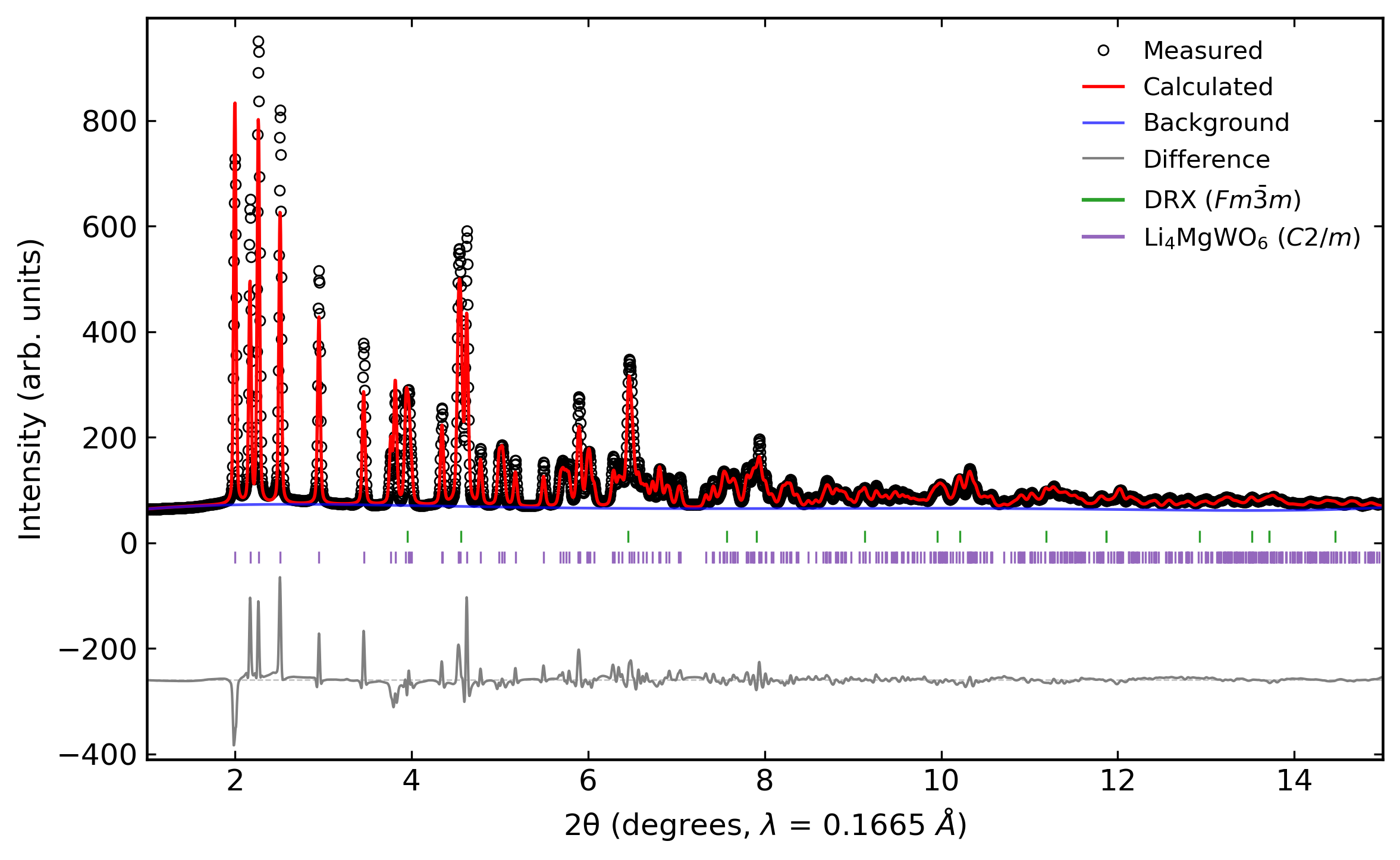}
\end{center}
\caption{Fit quality from a 2-phase refinement on synchrotron X-ray powder diffraction data collected on a disordered rocksalt cathode material with a nominal composition of Li$_{1.2}$Mn$_{0.4}$Mg$_{0.2}$W$_{0.2}$O$_2$. Weighted profile \textit{R}-factor (\textit{R}$_{wp}$) = 8.46\%.}
\label{fig:figure3}
\end{figure}

The fit profile (see Figure \ref{fig:figure3}) shows good agreement between the measured (black) and calculated (red) patterns with a reasonable background model (blue) and minimal difference (grey) across the full pattern, further supported by the weighted-profile \textit{R}-factor (\textit{R}$_{wp}$) of 8.46\%. PowderLine returned the quantitative descriptors of the sample, including the lattice parameters and peak broadening terms with ESDs for each phase, along with the the refined scale factors which were then used to calculate relative phase fractions (see Table \ref{tab:table1}). This demonstrates that a single recipe can drive a multi-phase refinement, with PowderLine returning structured outputs for direct use.

\begin{table}[H]
\centering
\caption{Refined structural and microstructural parameters with estimated standard deviations for the two crystalline phases fit in the synchrotron diffraction data on a disordered rocksalt cathode sample. Parameters extracted from a single two-phase Rietveld refinement executed with PowderLine. The weighted-profile $R$-factor for the refinement is $R_{\mathrm{wp}} = 8.46\%$.}
\label{tab:table1}
\vspace{0.5cm} 
\begin{tabular}{ccc}
\hline
Parameter & Phase 1 (DRX) & Phase 2 \\
          & Li$_{1.2}$Mn$_{0.4}$Mg$_{0.2}$W$_{0.2}$O$_2$ & Li$_4$MgWO$_6$ \\
          & Cubic ($Fm\bar{3}m$) & Monoclinic ($C2/m$) \\
\hline
$a$ (\AA)                & 4.1832(7)  & 5.1301(24)   \\
$b$ (\AA)                & ---        & 8.7881(5)  \\
$c$ (\AA)                & ---        & 5.1024(23)   \\
$\beta$ ($^{\circ}$)     & ---        & 110.791(6)   \\
Microstrain (\%)         & 0.0133(4)  & 0.0110(1)     \\
Phase fraction (wt.\%)   & 32.0(7)    & 68.0(7)     \\
\hline
\end{tabular}
\end{table}

\section{Requirments and performance}
\subsection{Machine and software requirements}
PowderLine runs on Linux and Windows, requires Python 3.10 or later, either a pinned build of GSAS-II or local TOPAS installation, and is installed as a reproducible software environment managed by Pixi \cite{fischer2025}. PowderLine and its direct dependencies have a memory footprint of approximately 80 MB. The refinement engine and execution add additional overhead (\textit{e.g.,} loading the GSAS-II engine adds about 70 MB, and a complete refinement peaks near 240 MB).

\subsection{Execution speed and server mode}
Reusing a refinement program through the persistent server mode (Section 2.4) has been seen to provide a significant increase in calculation speed (\textit{e.g.,} up to a tenfold speedup for single-phase fit with ~30 or less parameters), measured on a 16-core Xeon virtual machine with 23 GB of RAM. This speedup is largest for quick, simple refinements, where engine startup would otherwise dominate the total runtime, and it diminishes as the refinement itself becomes more expensive. Running the software as a long-lived service also allows refinement to be managed and scaled with standard process-orchestration tooling (\textit{e.g.,} Kubernetes). Each recipe is self-contained and returns its results in memory, so refinements remain independent of one another, and many patterns can be processed concurrently with throughput that scales with the number of available workers.

\section{Conclusions and outlook}

We developed PowderLine, a Python application that expresses a whole-pattern fit to powder diffraction data as a single declarative recipe, validates it against a versioned schema, and executes it through a refinement program to return structured results. The composability of PowderLine allows powder diffraction analysis to be done identically in varied environments, such as interactive work at a terminal, interactive notebooks, scripts, pipelines, and automated workflows. Through PowderLine's API coding agents are able to interface with refinement engines, bespoke graphical user interfaces can be rapidly developed, and custom analysis pipelines can be created while reusing common infrastructure.

At NSLS-II, we expect to use PowderLine extensively at Bluesky-enabled powder diffraction beamlines \cite{Allan2019}. For beamline operations PowderLine enables tools for automated instrument calibration and anomaly detection (\textit{e.g.,} instrument geometry has changed), as well as real-time data analysis via an API call. AI-driven workflows can thus be conditioned on reproducible diffraction results, and rich measurement metadata can be captured along with the sample provenance and analysis inputs/outputs. PowderLine can also be used as a simulation engine to generate training data for AI/ML models, and those models can then slot into full stack applications for automated analysis. PowderLine's open development invites community-contributed extensions to both its capabilities and supported refinement software, as well as integration with data-acquisition systems.

\begin{acknowledgements}
The authors thank Dr. Saul Lapidus and Dr. Zhuo Li for their valuable feedback. 
\end{acknowledgements}

\begin{funding}
This research was supported by DOE/Basic Energy Sciences under the Genesis Mission BES AI Pathfinder Program, Integrated Scientific Agentic AI for Catalysis (ISAAC). This research used resources of the National Synchrotron Light Source II, a U.S. Department of Energy (DOE) Office of Science User Facility operated for the DOE Office of Science by Brookhaven National Laboratory under Contract No. DE-SC0012704.\end{funding}

\ConflictsOfInterest{The authors declare no conflicts of interest.
}

\DataAvailability{PowderLine is open-source software released under the BSD-3-Clause license, and its source code, documentation, and worked examples are available at \url{https://github.com/NSLS2/PowderLine}. Hosted documentation is available at \url{https://powderline.readthedocs.io/}. The repository is publicly available, and questions, bug reports, and contributions are handled through GitHub or by contacting the corresponding author.}

\bibliography{powderline_v2}

\end{document}